\documentstyle[preprint,prd,aps]{revtex}
\begin{document}
\tightenlines 
\preprint{
\vbox{\halign{&##\hfil\cr
& ANL-HEP-PR-98-29\cr
& OHSTPY-HEP-T-98-008  \cr
& June, 1999    \cr
&\vspace{0.6truein}   \cr
}}}
   
\title{Renormalon Ambiguities in NRQCD Operator Matrix Elements}

\author{Geoffrey T.\ Bodwin} 
\address{High Energy Physics Division, Argonne National Laboratory, Argonne, IL
60439} 
\author{Yu-Qi Chen}
\address{Physics Department, Ohio State University, Columbus, Ohio 43210}

\maketitle

\begin{abstract}
We analyze the renormalon ambiguities that appear in factorization
formulas in QCD. Our analysis contains a simple argument that the
ambiguities in the short-distance coefficients and operator matrix
elements are artifacts of dimensional-regularization factorization
schemes and are absent in cutoff schemes. We also present a method for
computing the renormalon ambiguities in operator matrix elements and
apply it to a computation of the ambiguities in the matrix elements that
appear in the NRQCD factorization formulas for the annihilation decays
of S-wave quarkonia. Our results, combined with those of Braaten and
Chen for the short-distance coefficients \cite{braaten-chen}, provide an
explicit demonstration that the ambiguities cancel in the physical decay
rates. In addition, we analyze the renormalon ambiguities in the
Gremm-Kapustin relation and in various definitions of the heavy-quark
mass. 

\end{abstract}


\vfill \eject

\narrowtext 

\section{Introduction} 
\label{sec:introduction}

In Quantum Chromodynamics (QCD), it is often useful to describe physical
processes involving more than one distance scale by making use of a
factorization formalism. In such a formalism, a physical observable is
written as a sum of products of short-distance coefficients with
long-distance operator matrix elements. The short-distance coefficients
may be calculated as a perturbative series in the strong coupling
constant $\alpha_s$, evaluated at the short-distance scale. The
operator matrix elements contain all of the sensitivity of the physical
observable to low-momentum (infrared) processes. Because of this
infrared (IR) sensitivity, the operator matrix elements are generally
not amenable to a perturbative calculation and are usually determined by
comparison of physical quantities with experimental values or through
nonperturbative methods, such as lattice QCD. Some well-known examples
of this approach are the light-cone expansion in deep-inelastic
scattering, the factorization expressions for hard hadron-hadron cross
sections, Heavy-Quark Effective Theory (HQET), which is useful in the
study of heavy-light mesons, and Nonrelativistic Quantum Chromodynamics
(NRQCD), which is useful in the study of heavy quarkonium. 

The conventional wisdom is that the perturbation expansions for the
short-distance coefficients in QCD factorization formulas are, at best,
asymptotic series \cite{thooft}.  One can see why this might be the case
by examining the behavior of the perturbation series in the context of a
simple model, which gives the leading large-$N_f$ behavior of the
theory. This model has seen extensive applications in recent years in
such topics as the Operator Product Expansions (OPE) for $e^+e^-$
annihilation and $\tau$ decay\cite{Mueller,Beneke,Neubert,BBB}, Heavy
Quark Effective Theory (HQET)\cite{BB94,BSU,NS94,LMS94}, and the NRQCD
factorization formalism\cite{braaten-chen}. In this model, which we call
the ``bubble-chain model,'' one generates the ``perturbation series''
by inserting into the gluon propagator in a one-loop diagram all
numbers of order-$\alpha_s$ fermion-loop vacuum-polarization corrections.
The resulting perturbation series gives the order-$\alpha_s$
fermion-loop contribution to the running of the coupling constant. In
this expression, one then replaces the fermion-loop contribution to the
one-loop QCD beta function $\beta_0$ with the full one-loop QCD beta
function. As we will describe later, in the bubble-chain model, the
perturbation series diverges, with terms growing as the factorial of the
order. 

One can attempt to resum the series by carrying out a Borel
transformation. In the Borel plane, the factorial growth of the series
corresponds to singularities, known as renormalons. If the renormalon
singularities lie on the positive real axis, then they make the inverse
Borel transform ill-defined.  These ambiguities in the inverse Borel
transform are potentially important because they have the same nominal
size as corrections in the QCD factorization formulas that go as powers
of the ratio of scales. 

In a well-defined theory, physical observables, in contrast with
short-distance coefficients, are unambiguous. Of course, if one were to
attempt to compute a physical observable entirely in perturbation
theory, without making use of a factorization formalism, then one might
encounter ambiguities, owing to the failure of the perturbation series
to converge. However, such ambiguities are not intrinsic to the
observable. They are artifacts of the method of computation and would be
absent if one were to compute the observable in terms of nonperturbative
expressions, such as lattice path integrals. 

When one computes a physical observable by making use of a factorization
scheme that is defined in perturbation theory, such as one that is based
on dimensional regularization, then the short-distance coefficients and
operator matrix elements can contain ambiguities that arise in
consequence of the defining perturbation series. It is these ambiguities
that arise from the factorization scheme that are the focus of this
paper. 

If one were to attempt to compute an operator matrix element entirely in
perturbation theory, then additional ambiguities, which have nothing to
do with the factorization scheme, might arise. Such ambiguities are not
intrinsic to the matrix element, since they would not be present if one
defined the matrix element nonperturbatively and evaluated it in terms
of nonperturbative expressions. As we shall describe later in this
paper, one can isolate the ambiguities that arise from the factorization
scheme by computing the difference between an unambiguous definition of
the operator matrix element (for example, a lattice definition) and an
ambiguous definition (for example, one based on dimensional
regularization). In the remainder of this paper, when we refer to
ambiguities in operator matrix elements, we mean only those ambiguities
that arise from the factorization scheme. 

Because physical observables are unambiguous, the renormalon ambiguities
in the short-distance coefficients must be cancelled by corresponding
ambiguities in the operator matrix elements, to the level of accuracy of
the factorization procedure. Factorization procedures that are based
entirely on an underlying effective field theory, such as those in the
light-cone expansion, OPE, HQET, and NRQCD, are, in principal, valid to
all orders in an expansion in the inverse of the large scale in the
process. In contrast, factorization procedures for hadron-hadron-induced
hard-scattering processes, for example, the Drell-Yan process, are known
to fail for contributions that are subleading in the inverse of the
large scale \cite{doria-et-al}. In the discussions in the remainder of
this paper, we have in mind the factorization procedures that are based
entirely on underlying effective field theories. However, our results
may also apply to other factorization procedures at the level of leading
orders in the inverse of the large scale.  

If one determines operator matrix elements by comparing factorized
expressions for physical observables with results from experiment, then
the cancellation of renormalon ambiguities holds by construction, for
those observables. On the other hand, if one determines operator
matrix elements through a calculation in the underlying effective field
theory, as we do in this paper, then the cancellation of renormalon
ambiguities is a nontrivial confirmation of the accuracy of the
effective field theory in reproducing the low-momentum behavior of the
full theory and of the consistency of the definitions of the
short-distance coefficients and operator matrix elements. 

In this paper, we investigate the relations between the renormalon
ambiguities in the short-distance coefficients and the renormalon
ambiguities in the operator matrix elements.  In our analysis, it is
clear that the short-distance coefficients and the operator matrix
elements display no ambiguities in a factorization scheme in which
ultraviolet (UV) divergences in the operator matrix elements are
controlled with a regulator, such as lattice regularization, that has an
explicit cutoff in the loop momenta \cite{novikov-et-al,david}. In
contrast, in a dimensional-regularization scheme, both the
short-distance coefficients and the operator matrix elements contain
renormalon ambiguities, but they cancel in physical observables
\cite{LMS94}. We give a prescription for calculating the renormalon
ambiguities in the dimensionally-regulated operator matrix elements, and
we apply this method to the computation of the renormalon ambiguities in
the NRQCD matrix elements that appear in the decays of S-wave
quarkonia in the leading and first subleading orders in the expansion in
the heavy-quark velocity. Comparing with a calculation of the
ambiguities in the short-distance coefficients for the decays of S-wave 
quarkonia \cite{braaten-chen}, we find that the renormalon ambiguities
cancel in the physical decay rates, as expected. 

The remainder of this paper is organized as follows. In
Sec.~\ref{sec:renormalons}, we review the bubble-chain model and Borel
transformation and discuss, in general, the origins of renormalons in
short-distance coefficients. Here we present a concise argument to show
that renormalon ambiguities in the short-distance coefficients arise
from the low-momentum regions of loop integrals.
Sec.~\ref{sec:factorization} contains a discussion of the interplay
between the factorization scheme and the renormalon ambiguities. In
this section we clarify the fact that renormalon ambiguities in the
short-distance coefficients and operator matrix elements are artifacts
of dimensional-regularization factorization schemes and are absent in
cutoff schemes. In Sec.~\ref{sec:method}, we describe our method for
computing the renormalon ambiguities in the operator matrix elements. We
use this method, in Sec.~\ref{sec:computation}, to calculate the
renormalon ambiguities in the NRQCD matrix elements for decays of S-wave
quarkonia. In Sec.~\ref{sec:gremm-kapustin}, we examine the renormalon
content of the various quantities that appear in the Gremm-Kapustin
\cite{gremm-kapustin} relation. We also discuss the ambiguities in
various definitions of the heavy-quark mass. Finally, we summarize out
results and discuss their implications in Sec~\ref{sec:conclusion}. 

\section{Renormalons in the Bubble-Chain Model}
\label{sec:renormalons}

As we have mentioned, the bubble-chain model, gives the exact
large-$N_f$ behavior of QCD. In it, one generates the ``perturbation
series'' by inserting into the gluon propagator in a one-loop diagram
all numbers of order-$\alpha_s$ fermion-loop vacuum-polarization
corrections. That is, one replaces the factor $\alpha_s(\mu^2)$ in the
one-loop diagram with
\begin{mathletters}
\begin{eqnarray}
&&\alpha_s(\mu^2)\sum_{n=0}^\infty
\left[-\beta_0\alpha_s(\mu^2)\ln 
(l^2e^C/\mu^2)\right]^n\label{bubbles}\\
&&={\alpha_s(\mu^2)\over 1+\beta_0\alpha_s(\mu^2)\ln(l^2e^C/\mu^2)}
\label{bubble-sum}\\
&&\equiv \alpha_s^{(1)}(l^2),
\end{eqnarray}
\label{bubble-all}%
\end{mathletters}
where $l$ is the Euclidean momentum of the gluon in the one-loop
expression, $C$ is a renormalization-scheme-dependent constant ($C=-5/3$
in the $\overline{MS}$ scheme), $\beta_0$ is the one-loop
contribution to the QCD beta function, and we identify the expression
(\ref{bubble-all}) as the one-loop running coupling constant
$\alpha_s^{(1)}(l^2)$. At this stage, $\beta_0$ contains only the
fermion-loop contribution. However, in the bubble-chain model, one
includes the effects of gluons on the running of the coupling by
promoting $\beta_0$ to the complete QCD expression
$\beta_0=(33/2-N_f)/(6\pi)$. 

Now, suppose that a part of the original one-loop integral has the IR
behavior 
\begin{equation}
\int^\Lambda dl\, l^m.
\label{one-loop}
\end{equation}
Here we have introduced a cutoff $\Lambda$ on the magnitude of $l$, in
order to focus on the low-momentum region. We assume that the integral
is IR finite ($m\ge -1$), since any infrared divergences
are absorbed into the operator matrix elements in the factorization
procedure. If one inserts the expression (\ref{bubbles}) into the
one-loop integral (\ref{one-loop}) and integrates term by term, then it
is easy to see, using 
\begin{equation}
\int_0^\mu dl\, l^m \ln^n (l^2/\mu^2)=2^n
\mu^{m+1}(-1)^n n!/(m+1)^{n+1},
\end{equation}
that terms in the resulting series grow as $n!$.  Hence, the series is 
only asymptotic.  

One way to associate a well-defined function with an asymptotic series 
is through the Borel transform.  If a function $f(\alpha_s)$ has a 
power-series expansion
\begin{equation}
f(\alpha_s) \;=\;  \sum_{n=0}^{\infty}\, a_n \alpha_s^{n} \;,
\label{original}
\end{equation}
then its Borel transform is defined by
\begin{equation}
\widetilde{f}( t ) \;=\; a_0 \delta(t) \;+\; \sum_{n=1}^{\infty}
 { 1 \over (n-1)! }\;  a_{n} t^{n-1} \;.
\label{borel}
\end{equation}
In this paper, we indicate the Borel transform of a quantity $f$
either as $\tilde f$ or $B[f]$. Clearly, the series (\ref{borel}) for
the Borel transform has better convergence properties than the original
power series (\ref{original}). The function $f(\alpha_s)$ that generates
the series (\ref{original}) can be recovered from the Borel transform
$\widetilde{f}(t)$ through the inverse Borel transform 
\begin{equation} 
f(\alpha_s) \;=\; \int_0^{\infty} \;dt \; e ^{-t/\alpha_s} \; \widetilde{f} 
(t)\;,
\label{borel-inverse} 
\end{equation} 
provided that the integral (\ref{borel-inverse}) is sufficiently
convergent and that the Borel transform (\ref{borel}) is well-defined
along the positive real axis. \footnote{If the power series
(\ref{original}) is only asymptotic, then there is not a unique function
$f(\alpha_s)$ that generates it.  The inverse Borel transform picks out a
function with particular analyticity properties \cite{bender-orszag}.} 

The Borel transform of the series (\ref{bubbles}) is 
\begin{equation}
\widetilde{\alpha}_s^{(1)}(t)= {(\mu^2/e^C)^u\over l^{2u}},
\label{bubbles-tx}
\end{equation}
where $u=\beta_0t$.  If we insert the Borel transform (\ref{bubbles-tx})
into the one-loop integral (\ref{one-loop}), we obtain
\begin{equation}
(\mu^2/e^C)^u\int_0^\Lambda dl\, l^{m-2u}
=(\mu^2/e^C)^u{\Lambda^{m+1-2u}\over m+1-2u},
\label{bubble-integral}
\end{equation}
where $\Lambda$ is an ultraviolet cutoff, and, in the spirit of
dimensional regularization, we have defined the integral for ${\rm
Re}~u>(m+1)/2$ by analytic continuation. The integral has a pole in the
complex $u$ plane at $u=(m+1)/2$.  This singularity appears even though
the original one-loop integral is perfectly well-behaved.  A pole in
$u$, such as this one, that is associated with the running of $\alpha_s$
is called a renormalon\cite{thooft}. 

In order to understand the origins of this renormalon singularity more
fully, let us re-examine the integral (\ref{bubble-integral}).  Applying
an IR cutoff $\lambda$, we have 
\begin{equation}
(\mu^2/e^C)^u\int_\lambda^\Lambda dl\, l^{m-2u}
=(\mu^2/e^C)^u\;{\Lambda^{m+1-2u}-\lambda^{m+1-2u}\over m+1-2u}.
\label{cutoff-integral}
\end{equation}
If $\Lambda$ and $\lambda$ are finite positive numbers, then the
integral (\ref{cutoff-integral}) is well defined, even at the point
$u=(m+1)/2$.
Furthermore, the inverse Borel transform is convergent, provided that 
\begin{equation}
\alpha_s(\mu^2)\beta_0\ln[\mu^2/(\lambda^2e^C)]<1,
\label{pert-cond}
\end{equation}
which is identical to the condition that the integration over $l$ never
passes through the Landau pole in (\ref{bubble-sum}).  Hence, we
conclude that the renormalon singularity in the $u$~plane arises from
the region of integration near $l=0$, which we have excluded by
introducing the IR cutoff $\lambda$.  The renormalon singularity is a
signal that, because of the growth of the running coupling, the
perturbation series breaks down near zero loop momentum. A renormalon 
that arises from the region near zero loop momentum is called an IR 
renormalon.

The renormalon in (\ref{bubble-integral}) is on the positive real
axis. Hence, the inverse Borel transform is not unique.  One can
obtain various values, depending on the prescription used to deform the
integration contour near the pole. For example, one can deform the
integration contour in (\ref{borel-inverse}) into the complex plane so
that it runs above the pole or below the pole, or one can take a linear
combination of these two contours, such as the principal value. These
prescriptions all differ by amounts that are proportional to the residue
of the integrand of (\ref{borel-inverse}) at the pole. If the pole in
$\widetilde{f} (t)$ is at the point $t^*=u^*/\beta_0$ and has residue
$R^*$, the ambiguity in $f(\alpha_s)$ has the form 
\begin{equation}
\Delta f (\alpha_s) \;=\; K\, (2 \pi \,R^* )
      \, e^{- { t^* / \alpha_s(\mu^2)  } } \;,
\label{ambigu}
\end{equation}
where $K$ is a constant of order unity. In $\alpha_s(\mu^2)$, the scale
$\mu$ is typically chosen to be the largest scale in the physical
process. For large $\mu$ 
\begin{equation}
\alpha_s (\mu) \approx
{1 \over \beta_0 \ln ({\mu^2 / \Lambda_{\rm QCD} ^2 }) }.
\label{alphas}
\end{equation}
Inserting this expression into (\ref{ambigu}), we see that the
renormalon ambiguity is given approximately by
\begin{equation}
\Delta f \;\approx \; K ( 2 \pi \,R^*)\,\left({\Lambda_{\rm QCD} ^2
\over \mu^2 }  \right)^{u^*} \;.
\label{ambigu:2}
\end{equation}
Thus, the ambiguity is suppressed as a power of $\mu^2$. However,
such an ambiguity can be of practical importance, since it is of the
same nominal size as power corrections in the factorization formulas.

We have seen that IR renormalons can arise from the region of
integration near zero loop momentum. In an analogous fashion, UV
renormalons can arise from the region of integration near infinite loop
momentum. The loop integration extends to infinity in unregulated,
UV-finite integrals and in UV-divergent integrals in dimensional
regularization. In a renormalizable theory, the loop integration in a
short-distance coefficient is, at worst, logarithmically divergent in
the UV region when $u=0$.  (The integrand that we have displayed in the
expression (\ref{bubble-integral}) is an approximate form that is valid
only in the IR region.) In the case of a logarithmic UV divergence, the
pole at $u=0$ is identical to the dimensional-regularization pole in
$\epsilon=(4-D)/2$, where $D$ is the number of space-time dimensions,
and it is removed in the standard minimal-subtraction renormalization
procedure. The remaining pieces of the integrand, which have a
convergent power count in the large-momentum region when $u=0$, may
contribute additional renormalons.  However, by virtue of their
convergent power count, these pieces yield renormalons only on the
negative real $u$~axis. Renormalons on the negative real $u$~axis do not
introduce ambiguities into the inverse transform in
Eq.~(\ref{borel-inverse}). 

Another potential source of renormalons in the short-distance
coefficients is the UV renormalization procedure that is applied to full
QCD.  If the renormalization counterterms involve loop integrations down
to zero momentum, then the counterterms can contain renormalons. In this
paper, we assume that a minimal-subtraction procedure ($MS$) or modified
minimal-subtraction procedure ($\overline{MS}$) is used, so that the
renormalization counterterms are poles in $\epsilon$ or constants and
contain no IR renormalons. 
 
\section{Factorization-Scheme Dependence of Renormalons}
\label{sec:factorization}

In a factorization formalism, the domain of integration of a loop
integral in full QCD is partitioned between the operator matrix elements
and the short-distance coefficients.  This partitioning allocates the
low-momentum part of the loop integral, including the IR divergences,
to the operator matrix elements and allocates the remaining
high-momentum part of the loop integral to the short-distance
coefficients, which contain no IR divergences. 

The partitioning arises in the matching of the effective theory to full
QCD. For example, in the on-shell matching procedure, on-shell quark
and gluon amplitudes in the full theory are equated to the same
amplitudes in the effective theory. Each amplitude in the effective
theory can be written as a sum of products of short-distance
coefficients with matrix elements in the on-shell quark and gluon
states. Low-momentum parts of the amplitudes in the full theory,
including all of the IR divergences, correspond to matrix elements in
the effective theory. These matrix-element contributions can be factored
from the full amplitudes. Then, the remaining IR-finite parts of the
full amplitudes correspond to the coefficient functions. In the
perturbative implementation of the matching procedure, which is
familiar, for example, from treatments of the QCD-improved parton model,
the matrix-element contributions are subtracted from the full-QCD
amplitudes order-by-order in $\alpha_s$. Then, the remainders from this
subtraction procedure are identified with perturbative contributions to
the short-distance coefficients in the effective theory. 

The amount of the full-QCD amplitude that resides in the matrix elements
and, hence, in the short-distance coefficients, is controlled by a UV
regulator that is imposed on the operator matrix
elements.\footnote{Throughout this paper, we speak of the short-distance
coefficients as being dimensionally regulated, lattice regulated, etc.
What we mean is that these are the short-distance coefficients that
correspond, in the factorization formalism, to operator matrix elements
that are dimensionally regulated, lattice-regulated, etc. Of course, one
can apply an IR regulator to the short-distance coefficients as an
intermediate step in calculations. However, dependences on such IR
regulators cancel in the matching conditions, and the IR behavior of the
short-distance coefficients is controlled by the UV regulator of the
operator matrix element.} The UV cutoff plays the r\^ole of a
factorization scale, and the choice of UV regulator is known as the
``factorization scheme.''  The consistency of the underlying effective
field theory guarantees that physical quantities are independent of the
factorization scheme, provided that one works to sufficient accuracy in
the effective field theory. 

One class of UV regulator consists of those regulators in which the
cutoff is manifest and alters the form of the integrand at large
momentum to make it more convergent.  We call such regulators ``cutoff''
regulators.  Pauli-Villars regulators, lattice regulators, and explicit
bounds on the magnitude of the Euclidean momentum are all examples of
cutoff regulators. For such regulators, power UV divergences manifest
themselves as explicit power dependences of the matrix elements on the
UV cutoff. For our purposes, the crucial property of cutoff regulators
is that they do not alter the behavior of the integral below the
cutoff.\footnote{Lattice regulators and Pauli-Villars regulators {\it
do} alter the behavior of the integral below the cutoff, but only
through terms that scale as inverse powers of the cutoff. In an
effective field theory, one compensates for this effect, restoring the
behavior of the theory below the cutoff to the required level of
accuracy, by including in the effective action corresponding terms that
scale as inverse powers of the cutoff.}

In contrast, in dimensional regularization in $D$ dimensions, the cutoff
enters only implicitly through an overall factor that goes to unity when
$2\epsilon=4-D$ vanishes. After one discards poles in $\epsilon$, one
finds that logarithmic divergences {\it are} cut off by the
dimensional-regularization scale.\footnote{When we speak of
dimensionally-regulated matrix elements in this paper, we assume that
such poles, and possibly some associated constants, have been discarded.
That is, by dimensionally-regulated matrix elements, we really mean $MS$
or $\overline{MS}$ matrix elements.} However, terms with a
power-divergent power count that are homogeneous in the integration
momentum vanish.\footnote{We use the term ``homogenous'' to denote
expressions that are a pure power or a pure power times logarithms.} 
Because of this property, dimensional regularization potentially alters
the behavior of the integral below the cutoff. 

In a cutoff factorization scheme, the loop-integration for the IR-finite
parts of the short-distance coefficients does not extend to zero
momentum.  The part of the integration below the cutoff resides in the
operator matrix elements, to the extent that the effective
theory accurately reproduces the low-momentum behavior of the full
theory. Hence, in a cutoff factorization scheme, the short-distance
coefficients are free of a given IR renormalon ambiguity, provided that
the effective action retains enough powers of the small scale in the
calculation to reproduce the low-momentum behavior that is
characteristic of such a renormalon. The absence of renormalons in a
cutoff factorization scheme has been discussed in a number of
publications. Some of the earliest discussions are in
Refs.~\cite{novikov-et-al,david}. 

In a dimensional-regularization factorization scheme, it is conventional
to integrate the IR-finite parts of the short-distance coefficients down
to zero momentum.  Consequently, the short-distance coefficients contain
renormalon ambiguities. The IR-finite parts of the short-distance
coefficients correspond to the power-UV-divergent parts of the operator
matrix elements. Thus, because of the constraints imposed by the
matching conditions, the integration of the IR-finite parts of the
short-distance coefficients down to zero momentum demands, for
consistency, that the UV-power-divergent parts of the matrix elements be
set to zero. As we will explain in Sec.~\ref{sec:dim-reg-prescrip},
there is a prescription for regulating the operator matrix elements
dimensionally in which such power UV divergences {\it are} set to zero.
This prescription implies that the dimensionally-regulated matrix
elements do not completely reproduce the behavior of the full theory
below the factorization scale. 

Matrix elements are completely determined, in principle, in terms of
short-distance coefficients and physical observables. Therefore, the
absence of ambiguities in the cutoff short-distance coefficients
implies that the cutoff matrix elements are also ambiguity-free.
This is consistent with the fact that lattice-regulated matrix elements
have an unambiguous definition in terms of path integrals in the
effective theory. In a cutoff scheme, the low-momentum content of 
the theory resides completely in the operator matrix elements.

In a dimensional-regularization scheme, the short-distance coefficients
contain ambiguities.  Therefore, the dimensionally-regulated operator
matrix elements must contain ambiguities that cancel those in the
short-distance coefficients. The appearance of renormalon ambiguities
in the short-distance coefficients and operator matrix elements in the
dimensional-regularization scheme has no physical significance: It is a
factorization-scheme artifact. We conclude that the factorial growth in
the perturbation series for the short-distances coefficients is
cancelled by the factorial growth in the perturbation series that relate
cutoff-regulated matrix elements to dimensionally-regulated matrix
elements, provided that one works to the same order in $\alpha_s$ in
both quantities, and provided that one works to sufficient accuracy in
the underlying effective field theory \cite{LMS94}. 

Let us explain more fully what we mean by sufficient accuracy in the
underlying effective field theory. Generally, an operator matrix element
scales, in its leading behavior, as a power of the small scale in the
two-scale problem. For example, in the NRQCD factorization formalism for
heavy-quarkonium decay and production \cite{bbl}, matrix elements scale
as powers of $p=mv$, where $p$ and $v$ are the typical
heavy-quark-antiquark relative momentum and velocity, and $m$ is the
heavy-quark mass. To achieve the cancellation of a renormalon at given
value of $u$, it is crucial to retain matrix elements of sufficiently
high order in the small scale to reproduce the IR power behavior that
gives rise to that renormalon. Furthermore, if one is computing a
matrix-element renormalon ambiguity in the context of an effective field
theory, it is necessary to retain terms in the effective action of
sufficiently high accuracy in the small scale to reproduce the
corresponding IR power behavior. The action of an effective theory may
depend on both the cutoff and the large scale. For example, in HQET
and NRQCD, the action depends on $m$, as well as the cutoff. Hence,
the small scale may appear in the ratio of the small scale to the large
scale or in the ratio of the small scale to the cutoff. In such cases,
it is necessary to retain terms of sufficient accuracy in both the
expansion in inverse powers of the large scale and the expansion in
inverse powers of the cutoff.
\vfill\eject

\section{method for computing renormalon ambiguities in matrix elements}
\label{sec:method}

\subsection{Short-distance expression for the ambiguity}
\label{sec:short-distance}

To identify the renormalon ambiguities in a dimensionally regulated
matrix element, we make use of the fact that corresponding
cutoff-regulated matrix element is free of ambiguities. Then,
instead of computing the dimensionally-regulated matrix element itself,
we compute only the effects of a change of regularization scheme from
cutoff to dimensional.  

The effects of a change of regularization are contained in finite
renormalizations of the operators in the underlying effective theory. In
general, in order to work out these finite renormalizations, it is
necessary to identify all of the renormalization counterterms in the
effective theory and to compute their finite coefficients. However, at
the one-loop level, this amounts merely to computing the difference
between the dimensionally-regulated and cutoff amplitudes.  (In
both the dimensionally-regulated and cutoff amplitudes, we use the
dimensionally-regulated expression for the vacuum-polarization
insertions in the gluon propagator.)

Because the renormalization parts of an operator matrix element are
short-distance quantities, we can analyze them
perturbatively.\footnote{In fact, the dimensionally-regulated matrix
elements have only a perturbative definition, either in terms of
physical observables or in terms of matrix elements in some other
scheme, such as lattice regularization, that {\it does} have a nonperturbative
definition.}  Furthermore, the renormalization parts are independent of
the external states. Therefore, we are free to make a choice of the
external states that is convenient for a perturbative analysis, namely,
states consisting of on-shell elementary quanta (quarks, gluons, etc.). 

We note that, for purposes of computing the ambiguities in the
dimensionally-regulated matrix elements, any reference cutoff
regulator will do. At the one-loop level, a particularly convenient
choice of cutoff regulator is a simple cutoff on the magnitude of
the three-momentum of the gluon in the loop. While such a cutoff is not,
in general, consistent with gauge invariance, it is compatible with the
QED-like nature of the gauge invariance at one loop. With this choice,
we compute 
\begin{equation}
\langle {\cal K} \rangle_{\hbox{dim}}-\langle {\cal K} 
\rangle_{\hbox{cutoff}}
=\int_0^{\rm dim} d^4l\, I_{\cal K}-\int dl_0\, \int_0^{\lambda} 
d^3\mbox{\boldmath $l$} 
\, I_{\cal K}
=\int dl_0\, \int_{\lambda}^{\rm dim} d^3\mbox{\boldmath $l$} 
\, I_{\cal K},
\label{matrix-diff}
\end{equation}
where ${\cal K}$ is an operator, $I_{\cal K}$ is the integrand
corresponding to its matrix element between on-shell states in one-loop
perturbation theory, and ``dim'' and ``cutoff'' denote dimensional and
cutoff ultraviolet regularization, respectively. We write ``dim'' as the
upper limit of integration to indicate dimensional regularization of UV
divergences. $\lambda$ is a cutoff on the magnitude of the 3-momentum
{\boldmath $l$} and corresponds roughly to the IR cutoff $\lambda$
discussed in Sec.~\ref{sec:renormalons}. We assume that $\lambda$ is
sufficiently large that the condition (\ref{pert-cond}) is satisfied, so
that the integration on the right side of Eq.~(\ref{matrix-diff}) does
not pass through the Landau pole.

The expression (\ref{matrix-diff}), being a renormalization-scheme
dependence, is IR finite, as is apparent from the form on the right side
of the equation. However, it may happen that the separate terms in the
left and middle expressions in Eq.~(\ref{matrix-diff}) require an IR
regulator.  Then, one must, for consistency with the complete IR-finite
expression, impose the same IR regulator on both terms. 

The expression on the right side of Eq.~(\ref{matrix-diff}) is the basis
for our calculation of the renormalon ambiguities.  One could equally
well use the middle expression in Eq.~(\ref{matrix-diff}), taking
advantage of the fact, explained in Sec.~\ref{sec:dim-reg-prescrip},
that the dimensionally-regulated integral vanishes for IR-finite terms
in the integrand, so that one need compute only the cutoff
integral.\footnote{It is interesting to note that the cutoff integral
in the middle expression in Eq.~(\ref{matrix-diff}) contains IR
renormalons, whereas the right side of Eq.~(\ref{matrix-diff}) contains
UV renormalons. Thus, we see that UV renormalons can be replaced by IR
renormalons through the manipulation of quantities that vanish in
dimensional regularization.} However, we wish to emphasize the
insensitivity of the renormalon ambiguities to the low-energy physics
by working with the right side of Eq.~(\ref{matrix-diff}), which
contains no low-momentum contribution. With this computational 
procedure, the dimensionally-regulated matrix elements contain only UV 
renormalons.

\subsection{Dimensional-regularization prescription}
\label{sec:dim-reg-prescrip}

Let us now explain more precisely what we mean by dimensional
regularization of operator matrix elements. It is well-known that
dimensionally-regulated expressions can sometimes be ambiguous when the
integrand involves a limiting procedure. For example, the result for a
dimensionally-regulated integral can be changed by carrying out formal
manipulations involving power series in the integration variable.
Consequently, it is necessary to specify how the integrand is to be
arranged before the dimensional regularization is imposed. 

In computing the short-distance coefficients in the
dimensional-regularization scheme, it is conventional to integrate
IR-finite terms down to zero loop momentum. As we have discussed in
Sec.~\ref{sec:factorization}, such an approach corresponds to setting to
zero the UV power divergences in the operator matrix elements.
Therefore, for consistency with conventional calculations of the
short-distance coefficients, we wish to follow a procedure for the
dimensional regularization of the operator matrix elements in which UV
power divergences are set to zero. Such a procedure is usually employed
in regulating, for example, the matrix elements for parton distributions
\cite{collins-soper}. 

One applies this dimensional regularization procedure to QCD corrections
to the matrix elements of operators between on-shell quark and gluon
states. Such expressions appear, for example, in
Eq.~(\ref{matrix-diff}), and also in the matching conditions between
full QCD and the effective field theory that fix the short-distance
coefficients. One first decomposes the QCD corrections to the operator
matrix element into a linear combination of the tree-level operators in
the effective theory. (In the case of NRQCD, this amounts to Taylor
expanding the integrand with respect to the external momenta and taking
appropriate linear combinations to form quantities of definite orbital
angular momentum, spin, and color.) All of the dependence on the
external momenta now resides in the tree-level operators. The
coefficients of the tree-level operators contain the integration over
the loop momentum $l$ and are now independent of the external momenta.
In some effective theories, such as NRQCD, the coefficients of the bare
operators may not be homogeneous in $l$.  In that case, one further
expands the integrand in a power series in $l$ divided by the large
scale ($m$ in the case of NRQCD). Then, each term in the series is
homogeneous in $l$. Now, any terms that are IR finite (that is, UV power
divergent) require no IR regulator. Hence, such terms are scale
invariant, and their integrals over all $l$ vanish in dimensional
regularization. We note that this formal procedure is equivalent to
absorbing the UV-power divergences into a renormalization of the
operator matrix elements. 

In light of these general considerations, we see that the application of
Eq.~(\ref{matrix-diff}) requires some care. To obtain a result that is
consistent with conventional calculations of the short-distance
coefficients in the dimensional-regularization scheme, it is essential
to use a procedure for computing the dimensionally-regulated matrix
elements, such as the one that we have specified, in which the UV power
divergences vanish. 

\section{NRQCD factorization formalism} 
\label{sec:nrqcd}

In order to establish our notation and conventions, we now give a brief
review of the NRQCD factorization formalism for heavy-quarkonium decays.

In the nonrelativistic limit, in which the heavy-quark (or antiquark)
velocity $v$ is much less than unity, full QCD is described accurately
by the effective theory NRQCD.  In particular, NRQCD is convenient for
describing heavy-quarkonium bound states.  In such systems, there are
three important energy scales: the heavy-quark mass $\sim m$, the
heavy-quark 3-momentum $\sim mv$, and the heavy-quark energy $\sim
mv^2$. The NRQCD effective theory is constructed from full QCD by
integrating out the effects from energies of order $m$ or larger. Since
NRQCD describes only the effects at energy scales less than $m$, there
is no heavy-quark-antiquark pair creation in the effective theory.  That
implies, to first approximation, that the heavy-quark and
heavy-antiquark fields are decoupled.  Hence, it is convenient to
express the effective action in terms of Pauli two-component
spinors---one for the heavy-quark field and one for the heavy-antiquark
field. 
 
The NRQCD effective Lagrangian can be used to reproduce the full theory
to any desired accuracy in $v$. We can write it, up to terms of relative 
accuracy $v^3$, as \cite{C-L,lepage-et-al} 
\begin{equation} 
{\cal L}_{\rm NRQCD} 
\;=\; {\cal L}_{0}\;+\; \delta {\cal L} \;, 
\end{equation} 
where 
\begin{equation} 
{\cal L}_{0} 
\;=\;  \psi^\dagger \left( iD_0 + {{\bf D}^2 \over 2m} \right)\psi
\;+\; \chi^\dagger \left( iD_0 - {{\bf D}^2 \over 2m} \right) \chi
\label{L_0}
\end{equation}
is the leading-order NRQCD effective Lagrangian, and  
\begin{eqnarray}
\delta{\cal L}
&=& {c_1 \over 8m^3} [\psi^\dagger ({\bf D}^2)^2 \psi
-\chi^\dagger ({\bf D}^2)^2 \chi]\nonumber\\
&&+{c_2 \over 8m^2}
[\psi^\dagger ({\bf D} \cdot g {\bf E} 
- g {\bf E} \cdot {\bf D}) \psi
+\chi^\dagger ({\bf D} \cdot g {\bf E} 
- g {\bf E} \cdot {\bf D}) \chi]
\nonumber \\
&&+{c_3 \over 8 m^2}
[\psi^\dagger (i {\bf D} \times g {\bf E}
 - g {\bf E} \times i {\bf D}) 
\cdot \mbox{\boldmath $\sigma$}\psi
+\chi^\dagger (i {\bf D} \times g {\bf E}
 - g {\bf E} \times i {\bf D}) 
\cdot \mbox{\boldmath $\sigma$}\chi]
\nonumber\\
&&+{c_4 \over 2m}
[\psi^\dagger (g {\bf B} \cdot \mbox{\boldmath $\sigma$}) \psi
-\chi^\dagger (g {\bf B} \cdot \mbox{\boldmath $\sigma$}) \chi]
\label{deltaL}
\end{eqnarray}
contains the relative order-$v^3$ corrections. Here,
$D^\mu=\partial^\mu+igA^\mu$, $\psi$ is the two-component Pauli field
that destroys a heavy quark, $\chi$ is the two-component Pauli field
that creates a heavy antiquark, and $c_1$, $c_2$, $c_3$, and $c_4$ are
dimensionless short-distance coefficients. They are determined by
matching amplitudes in the effective theory with those in full QCD. At
tree level, $c_1=c_2=c_3=c_4=1$. There is a spin symmetry in the lowest
order Lagrangian (\ref{L_0}), since it is independent of the spin of the
heavy quark and heavy antiquark. This symmetry is violated by the terms
in the higher-order corrections (\ref{deltaL}) that contain
$\mbox{\boldmath $\sigma$}$. 

In heavy-quarkonium annihilation decays, several distance scales are
involved. The annihilation of the heavy quark and antiquark occurs at
the short-distance scale $1/m$, whereas the dynamics of quark-antiquark
binding involves, principally, the long-distance scales $1/mv$ and
$1/mv^2$. The NRQCD factorization formalism\cite{bbl} separates the
physical effects that occur at the scale $1/m$ from those at the
longer-distance scales. Since effects from energies of order $m$ are
integrated out in obtaining the effective theory, the details of the
annihilation are not described in NRQCD. However, the amplitude for a
heavy quark and antiquark first to annihilate and then to be recreated
is taken into account through effective local four-fermion interactions.
The coefficient of a four-fermion operator contains the information
about the effects at the scale $m$; the matrix element of the
four-fermion operator in a quarkonium state contains the information
about the long-distance effects. 

The imaginary parts of the four-fermion interactions are related, 
through the optical theorem, to the total decay rate.  Hence, 
the heavy-quarkonium annihilation decay rate can be written in the 
factored form \cite{bbl} 
\begin{equation}
\Gamma ( H ) \;=\; {1 \over 2 M_{H} } \; 
\sum_{mn} {C_{mn} } \langle H | {\cal O}_{mn} | H \rangle \;,
\label{fact-Gam}
\end{equation}
where $C_{mn}$ is a short-distance coefficient that is proportional to
the imaginary part of the coefficient, in the effective action, of the
four-fermion operator, and $M_H$ is the  mass of the state $H$. The
matrix elements $ \langle H | {\cal O}_{mn} | H \rangle $ are
expectation values of local 4-fermion operators in the quarkonium state
$H$. These local operators have the general structure 
\begin{equation}
{\cal O}_{mn}
\;=\; \psi^\dagger {\cal K}_m \chi \;
	\chi^\dagger {\cal K}_n \psi \;,	
\label{O_mn}
\end{equation}
where ${\cal K}_n$ and ${\cal K}_m$ are direct products of a color
matrix ($1$ or $T^a$), a spin matrix ($1$ or $\sigma^i$), and a
polynomial in the gauge-covariant derivative ${\bf D}$ and
in the field strengths ${\bf E}$ and ${\bf B}$. 

If ${\cal K}_m$ and ${\cal K}_n$ are color-singlet operators, one can
use the vacuum-saturation approximation \cite{bbl} to simplify the
matrix elements.  This approximation is obtained by inserting  a
complete set of intermediate states between the quark-antiquark
bilinears and retaining only the vacuum hadronic state. The result is 
\begin{equation} 
\langle H|{\cal O}_{mn}| H \rangle\;\approx\;  
\langle H| \psi^\dagger {\cal K}_m \chi \;| 0 \rangle \langle 0| \; 
 	\chi^\dagger {\cal K}_n \psi | H \rangle \;. 
\label{M-E-factor} 
\end{equation} 
Hence, a matrix element can be factored into a product of the matrix
elements $\langle H| \psi^\dagger {\cal K}_m \chi \;| 0 \rangle $ and
$\langle 0| \; \chi^\dagger {\cal K}_n \psi | H \rangle $. In the case
of hadronic decays, this approximation is valid up to corrections of
relative order $v^4$ \cite{bbl}.  For electromagnetic decays, the
insertion of a vacuum projection operator is exact, since there are no
hadrons in the final state. 

Generally, the NRQCD matrix elements are nonperturbative in nature.
However, it is important to realize that they are fully determined, in
principal, by the NRQCD effective theory.  In fact, they can be computed
by making use of the lattice formulation of NRQCD \cite{B-S-K}.

\section{Computation of the $u=1/2$ renormalon in NRQCD matrix elements} 
\label{sec:computation}

In this section, we apply the method described in Sec.~\ref{sec:method}
to compute the $u=1/2$ renormalons, which yield the ambiguities that are
leading in $\Lambda_{\rm QCD}/\mu$, for the hadron-to-vacuum matrix
elements 
$\langle 0 |\chi^\dagger \sigma^i \psi | J/\psi \rangle  $, 
$\langle 0 |\chi^\dagger  \psi | \eta_c\rangle$,  
$\langle 0 |\chi^\dagger \sigma^i {\bf D}^2 
\psi |  J/\psi \rangle /m^2  $, 
$\langle 0 |\chi^\dagger {\bf D}^2  \psi |\eta_c \rangle /m^2 $.  
These are the matrix elements that arise in leading and next-to-leading
order in $v^2$ in the decays of the S-wave charmonium states. Following
Ref.~\cite{braaten-chen}, we employ a shorthand notation for these
matrix elements: 
\begin{mathletters}
\begin{eqnarray}
\langle  {\cal K} \rangle_{\psi} &\;\equiv \;&  
 \mbox{\boldmath $\epsilon$}\cdot 
 \langle 0 | \chi^\dagger \mbox{\boldmath $\sigma$}  \psi 
  |J/\psi(\mbox{\boldmath $\epsilon$})  \rangle,  \\
\langle  {\cal K} \rangle_{\eta} &\;\equiv \;&  
  \langle 0 | \chi^\dagger  \psi | \eta_c \rangle,  \\
 \langle {\cal K}_{ D^2} \rangle_{\psi} &\;\equiv \;& (1/ m^2) \,  
  \mbox{\boldmath $\epsilon$}\cdot \langle 0 | \chi^\dagger \mbox{\boldmath
  $\sigma$}{\bf D}^2  \psi | J/\psi(\mbox{\boldmath $\epsilon$}) 
\rangle , \\
 \langle {\cal K}_{ D^2} \rangle_{\eta} &\;\equiv \;& (1/m^2) \,  
   \langle 0 | \chi^\dagger {\bf D}^2  \psi | \eta_c \rangle \;.
\end{eqnarray}%
\end{mathletters}

As we have discussed in Section~\ref{sec:method}, the renormalon
ambiguity of the NRQCD matrix elements [Eq.~(\ref{matrix-diff})] is
insensitive to the low-energy behavior of the theory. It is, therefore,
independent of the external states. Thus, we can evaluate the operator
matrix elements in free $c\bar{c}$ states, using perturbative NRQCD
(pNRQCD).  That is, we use pNRQCD to evaluate the matrix elements 
\begin{mathletters}
\begin{eqnarray}
\langle  {\cal K} \rangle_{V} &\;\equiv \;&  
 \mbox{\boldmath $\epsilon$}\cdot 
 \langle 0 | \chi^\dagger \mbox{\boldmath $\sigma$}  \psi 
  |c\bar{c} (\mbox{\boldmath $\epsilon$})  \rangle,  \\
\langle  {\cal K} \rangle_{P} &\;\equiv \;&  
  \langle 0 | \chi^\dagger  \psi | c\bar{c} \rangle,  \\
 \langle {\cal K}_{ D^2} \rangle_{V} &\;\equiv \;& (1/m^2) \,  
  \mbox{\boldmath $\epsilon$}\cdot \langle 0 | \chi^\dagger \mbox{\boldmath
  $\sigma$}{\bf D}^2  \psi |  
c\bar{c}(\mbox{\boldmath $\epsilon$}) \rangle ,
  \\
 \langle {\cal K}_{ D^2} \rangle_{P} &\;\equiv \;& (1/m^2) \,  
   \langle 0 | \chi^\dagger {\bf D}^2  \psi | c\bar{c} \rangle \;.
\end{eqnarray}
\label{quark-matrix}%
\end{mathletters}
In each state, we take the quark and antiquark to be at rest and on 
their mass shells.  The latter choice makes the calculation manifestly 
gauge invariant.

\subsection{Feynman rules}
\label{sec:rules}

In the pNRQCD calculation, we take for the unperturbed Lagrangian the
noninteracting part of leading-order-in-$v$ NRQCD Lagrangian
(\ref{L_0}).  The interactions in Eq.~(\ref{L_0}), as well as the
higher-order-in-$v$ corrections in Eq.~(\ref{deltaL}) are treated as
perturbations.  The Feynman rules may be derived by standard methods.
The rules for the quark and antiquark propagators and the ``Abelian''
parts of the quark-gluon interaction vertices are shown in
Table~\ref{table:rules}. By ``Abelian'' we mean the terms that contain
no color-matrix commutators.  For our computation, which is at the
one-loop level, the non-Abelian terms do not contribute.

We also need the rules for the gluon propagators with all numbers of
fermion-loop vacuum polarization insertions. Each renormalized fermion-loop
vacuum-polarization yields a factor 
\begin{equation} 
 i\beta_0 \,  \alpha_s  \, 
\left[ \;\ln \left( {-l^2 -i\epsilon \over \mu^2 } \right)  \, + \, C \; 
\right] \, 
\,(- l^2 g^{\mu\nu} \,+\, l^\mu l^\nu)\;,
\label{bubble-prop}
\end{equation}
where $l$ is the gluon momentum. (As we have already mentioned,
$\beta_0$ contains only the fermion-loop contribution at this stage, but
eventually we promote $\beta_0$ to the complete QCD expression
$\beta_0=(33/2-N_f)/(6\pi)$.) For each vacuum-polarization insertion in
the gluon propagator, there is a factor (\ref{bubble-prop}) and a factor
from an additional free-gluon propagator. The sum over all numbers of
such vacuum-polarization insertions is a geometric series, which is
easily computed for the gluon propagator in a given gauge. In a
covariant gauge, the propagator with all numbers of vacuum-polarization
insertions is 
\begin{mathletters}
\begin{equation}
iD_{\rm cov}^{\mu\nu}=
\sum_{n=0}^\infty \beta_0^n \,  \alpha_s^n  \, 
\left[ \;-\ln \left( {-\l^2 -i\epsilon \over \mu^2 } \right)  \, - \, C \; 
\right]^n \, 
\,{ i(-g^{\mu\nu} \,+\, l^\mu l^\nu/l^2) \,\over l^2 +i\epsilon}  \;
-\, \xi \, {il^\mu l^\nu/l^2 \over l^2 + i \epsilon}\;,
\end{equation}
where we have suppressed color indices. Similarly, in the Coulomb gauge, 
the Coulomb-gluon propagator is
\begin{equation}
iD_C=
\sum_{n=0}^\infty \beta_0^n \,  \alpha_s^n  \, 
\left[ \;-\ln \left( {-l^2 -i\epsilon \over \mu^2 } \right)  \, - \, C \; 
\right]^n \, 
\,{ i\over \mbox{\boldmath $l$}^2 }\;,
\end{equation}
and the transverse-gluon propagator is
\begin{equation}
iD_T^{ij}=
\sum_{n=0}^\infty \beta_0^n \,  \alpha_s^n  \, 
\left[ \;-\ln \left( {-l^2 -i\epsilon \over \mu^2 } \right)  \, - \, C \; 
\right]^n \, 
\,{ i(\delta^{ij} \,-\, l^i l^j/\mbox{\boldmath $l$}^2) \over l^2 +i\epsilon }
\;.
\end{equation}
\label{complete-props}%
\end{mathletters}
In each instance, the $\alpha_s^0$ term is the free-gluon propagator.
Here, and throughout this paper, we use Greek letters for
Minkowski-space indices and Roman letters for 3-space indices. We do not
distinguish between upper and lower 3-space indices; both correspond to
the upper Minkowski-space index in covariant quantities and the lower
Minkowski-space in contravariant quantities. 

It is now straightforward to obtain the Borel transforms of $\alpha_s$
times the propagators in Eqs.~(\ref{complete-props}).  They are 
\begin{mathletters}
\begin{equation}
iB[\alpha_s D_{\rm cov}^{\mu\nu}]=
   \left( { \mu^2 \over e^C } \right)^u \;
    {  i(-g ^{\mu\nu} + l^\mu l^\nu/l^2)  \over (-l^2 -i\epsilon)^{1 + u} }\, 
    -\, \xi \, {il^\mu l^\nu/l^2 \over -l^2 - i \epsilon}
    \;,
\label{gauge}
\end{equation}
\begin{equation}
iB[\alpha_s D_C]=
   \left( { \mu^2 \over e^C } \right)^u\;
    {i  \over \mbox{\boldmath $l$}^2 \, (-l^2 - i\epsilon)^{ u} }  
    \;,
\label{gauge:longi}
\end{equation}
\begin{equation}
iB[\alpha_s D_T^{ij}]=
   - \, \left( { \mu^2 \over e^C } \right)^u\;
    {i(\delta_{ij} \,-\, l_i l_j/\mbox{\boldmath $l$}^2)  
    \over ( -l^2 - i\epsilon)^{1+ u} }  
    \;.
\label{gauge:trans}%
\end{equation}
\label{borel-complete-props}%
\end{mathletters}

\subsection{The computation}

Now we apply the method of Sec.~\ref{sec:method} to calculate the
$u=1/2$ ultraviolet renormalons associated with the matrix elements in
Eq.~(\ref{quark-matrix}).  For each perturbative correction to these
matrix elements, we compute the right side of Eq.~(\ref{matrix-diff}),
using the convention of Sec.~\ref{sec:dim-reg-prescrip} for the form of
the integrand in dimensional regularization. We consider only the
ambiguities that are proportional to the lowest-order matrix elements
$\langle {\cal K} \rangle_{V}$ and $\langle {\cal K} \rangle_{P}$. To
obtain the mixing into these matrix elements, we set the external
momentum equal to zero in perturbative corrections. We find it most
convenient to carry out the calculation in the Coulomb gauge. 

First we identify the $u=1/2 $ ultraviolet renormalons associated with
the matrix elements $\langle {\cal K}_{D^2} \rangle_{V}$ and $\langle
{\cal K}_{D^2}\rangle_{P}$.  Consider the vertex corrections to these
matrix elements that arise from a Coulomb-gluon exchange with two $A_0$
vertices.  These corrections contain pieces proportional to the
lower-order matrix elements $\langle  {\cal K} \rangle_{V}$ and
$\langle {\cal K} \rangle_{P}$, respectively, which are obtained by
setting the external momentum equal to zero. Since the $A_0$ vertices
respect the NRQCD spin symmetry, the vector and pseudoscalar
renormalization constants of proportionality are equal. They are given 
by
\begin{eqnarray} 
\delta\tilde{Z}_{V,D^2}=\delta\tilde{Z}_{P,D^2}
=4\pi i C_F \left( { \mu^2 \over e^C } \right)^u\; 
\int_\lambda^{\rm dim} {d ^4 l \over (2\pi)^4} \;
&&{1 \over l_0 - \mbox{\boldmath $l$}^2/(2m) + i \epsilon } \;
{\mbox{\boldmath $l$}^2 \over m^2 } \;
{1 \over -l_0 - \mbox{\boldmath $l$}^2/(2m) + i \epsilon } \;\nonumber\\
&&\times {1  \over \mbox{\boldmath $l$}^2\, (-l^2 - i\epsilon)^{ u} }  \;,
\label{D^2} 
\end{eqnarray} 
where $\int_\lambda^{\rm dim} d^4l$ is shorthand for $\int dl_0
\int_\lambda^{\rm dim} d^3\mbox{\boldmath $l$}$, and $C_F=4/3$. To
evaluate this integral, we first integrate over $l_0$, using the
contour-integral method. It is easy to see that the cut contribution has
a power count such that it cannot contribute a pole at $u=1/2$. The
contribution from the quark (or antiquark) pole yields 
\begin{equation}
\delta\tilde{Z}_{V,D^2}=\delta\tilde{Z}_{P,D^2}\sim 
-{16\pi \over 3 m}  \left( { \mu^2 \over e^C } \right)^u\; 
\int_\lambda^{\rm dim} {d^3 \mbox{\boldmath $l$} \over (2\pi)^3} \;
    {1  \over  \mbox{\boldmath $l$}^{2+2u}\, 
[1 - \mbox{\boldmath $l$}^2/(4m^2)]^u }\,  \;. 
\label{D^2:3q} 
\end{equation}
Now we expand the factor $[1-\mbox{\boldmath $l$}^2/(4m^2)]^{-1}$ in a
power series in $\mbox{\boldmath $l$}^2/m^2$. Only the first term in the
series has the correct power count to yield a pole at $u=1/2$.
Integrating over $\mbox{\boldmath $l$}$, we find the $u=1/2$ renormalon:
\begin{equation}
\delta\tilde{Z}_{V,D^2}=\delta\tilde{Z}_{P,D^2}\sim
{8  \over 3 \pi m } \;  
\left( { \mu^2 \over e^C } \right)^u \;
{1 \over 1-2 u} \;,
\label{D^2:u=1/2} 
\end{equation}
where we have used the fact that the contribution from the upper limit 
of the integration vanishes, since the integral is UV convergent in 
dimensional regularization.
We note that diagrams involving transverse gluons, higher-order NRQCD
vertices or the gauge fields in the ${\bf D}^2$ operator all contribute
with the wrong power count to yield a pole at $u=1/2$.  (In fact,
contributions involving transverse gluons and the part of an
$\mbox{\boldmath $\nabla$}\cdot {\bf A}_i$~vertex that is proportional
to $l$ vanish in the Coulomb gauge.) From (\ref{ambigu:2}), we obtain
the $u=1/2$ renormalon ambiguities in the NRQCD matrix elements:
\begin{mathletters}
\begin{eqnarray}
\Delta \langle {\cal K}_{ D^2} \rangle_{\psi} &\;=\;& -K{8 \over 3 \beta_0 }
 {\Lambda_{\rm QCD} \over e^{C/2} m} \langle {\cal K} \rangle_{\psi}  
\;,\\
\Delta \langle {\cal K}_{ D^2} \rangle_{\eta} &\;=\;& -K{8 \over 3 \beta_0 }
 {\Lambda_{\rm QCD} \over e^{C/2} m} \langle {\cal K}\rangle_{\eta}  \; .
\end{eqnarray}
\label{D^2-ambig}%
\end{mathletters}

Next we identify the $u=1/2 $ ultraviolet renormalons in the matrix
elements $\langle {\cal K} \rangle_{V}$ and $\langle {\cal K}
\rangle_{P}$ associated with the mixing of these matrix elements into
themselves. 

First consider the heavy-quark and heavy-antiquark self-energy diagrams.
It can be seen, by counting powers of momentum, that the diagrams
involving a transverse gluon or higher-order NRQCD vertices do not yield
a pole at $u=1/2$. The diagrams involving a Coulomb gluon contribute to
the $u=1/2$ renormalon through the renormalization constant of the quark
(antiquark) wavefunction. For a heavy quark with external energy $E$ and
momentum $p$, the Borel transform of the self-energy diagram is given by
\begin{equation}
\tilde \Sigma (E,{\bf p}) = 4 \pi i C_F \int 
{d^4 l \over (2\pi)^4} \;
{1 \over E+l_0 -({\bf p} - \mbox{\boldmath $l$})^2/2m + i\epsilon } \;
    {1  \over \mbox{\boldmath $l$}^2\, (-l^2 - i\epsilon)^{ u} }  \;.
\label{self:Sigma} 
\end{equation} 
The Borel transform of the renormalization constant of the 
wavefunction is 
\begin{equation}
\delta\tilde{Z}_2= {\partial\tilde \Sigma (E,{\bf p}) 
\over  \partial 
E}\biggr |_{E={\bf p}=0} =
-4 \pi i C_F \int {d^4 l \over (2\pi)^4} \;
{1 \over (l_0 -\mbox{\boldmath $l$}^2/2m + i\epsilon)^2 } \;
    {1  \over  \mbox{\boldmath $l$}^2 \, (-l^2 - i\epsilon)^{ u} }  \;.
\label{self:deltaZ}  
\end{equation} 
The antiquark self-energy differs from the quark self-energy only in the
sign of the $l_0 $ term in the quark (antiquark) propagator.  Hence,
these contributions are equal under a change of integration variables
$l_0\rightarrow -l_0$.  It is useful to symmetrize the integrand under
this change of variables. Then, the sum of the contributions
of the quark and antiquark wavefunction renormalizations yields
the renormalization coefficient
\begin{eqnarray} 
\delta\tilde{Z}_{\rm wf}\;=\;-4\pi i C_F \left( { \mu^2 \over e^C } \right)^u\; 
\int_\lambda^{\rm dim} {d ^4 l \over (2\pi)^4} \;
 &&\left[ \; 
{1 \over (l_0 - \mbox{\boldmath $l$}^2/(2m) + i \epsilon )^2} \;
 \;+\; 
{1 \over (-l_0 - \mbox{\boldmath $l$}^2/(2m) + i \epsilon )^2} \;
\right] \;\nonumber\\
&& \times
{1  \over  \mbox{\boldmath $l$}^2 \, (-l^2 - i\epsilon)^{ u} }  \;.
\label{self} 
\end{eqnarray} 
If we carry out the $l_0$ contribution by contour integration, the 
individual cut contributions yield poles at $u=1/2$.  However, it is 
easy to see that the sum of the cut contributions from the quark and 
antiquark terms does not have a pole at $u=1/2$.  Computing the double-pole 
contribution and integrating over $\mbox{\boldmath $l$}$, we obtain
\begin{equation}
\delta\tilde{Z}_{\rm wf} \sim  {2 \over 3 \pi m} \;  
\left( { \mu^2 \over e^C } \right)^u \;
{1 \over 1-2 u} \;.
\label{self:u=1/2} 
\end{equation}

Next we consider the vertex correction involving a Coulomb gluon with
$A_0$~vertices.  From the right side of Eq.~(\ref{matrix-diff}), we find 
that this vertex correction yields the renormalization coefficients
\begin{eqnarray} 
\delta\tilde{Z}_{V,a}=\delta\tilde{Z}_{P,a}
=-4\pi i C_F \left( { \mu^2 \over e^C } \right)^u\; 
\int_\lambda^{\rm dim} {d^4 l \over (2\pi)^4}\; 
&&{1 \over l_0 - \mbox{\boldmath $l$}^2/(2m) + i \epsilon } \;
{1 \over -l_0 - \mbox{\boldmath $l$}^2/(2m) + i \epsilon }\nonumber \\
&&\times    {1  \over  \mbox{\boldmath $l$}^2\, (-l^2 - i\epsilon)^{ u} }  \;.
\label{vertex}
\end{eqnarray}
If we carry out the $l_0$ integration by the contour method, we see that 
that cut contribution does not contain a pole at $u=1/2$.  However, the 
contribution of the quark (or antiquark) pole is 
\begin{equation}
\delta\tilde{Z}_{V,a}=\delta\tilde{Z}_{P,a}
\sim {16\pi m \over 3}  \left( { \mu^2 \over e^C } \right)^u\; 
\int_\lambda^{\rm dim} {d^3 \mbox{\boldmath $l$ } \over (2\pi)^3} \;
    {1  \over  \mbox{\boldmath $l$}^4 }{1\over [\mbox{\boldmath $l$}^2
-\mbox{\boldmath $l$}^4/(4m^2)]^u}.
\label{vertex:3q} 
\end{equation}
Expanding the last denominator in a power series in $\mbox{\boldmath
$l$}^2/m^2$, we find that the leading term contains a pole at $u=-1/2$,
but not at $u=1/2$. The second term in the power-series expansion of the
last denominator in Eq.~(\ref{vertex:3q}) does yield a pole at $u=1/2$: 
\begin{equation}
\delta\tilde{Z}_{V,a}=\delta\tilde{Z}_{P,a}
\sim {4u\over 3\pi m}  \left( { \mu^2 \over e^C } \right)^u\; 
\int_\lambda^{\rm dim} {d^3 \mbox{\boldmath $l$} \over (2\pi)^3} \;
    {1  \over  \mbox{\boldmath $l$}^{2+2u} }.  
\end{equation}
Carrying out the integration over $\mbox{\boldmath $l$}$, we obtain
\begin{equation}
\delta\tilde{Z}_{V,a}=\delta\tilde{Z}_{P,a}
\sim -{1\over 3\pi m} \;  
\left( { \mu^2 \over e^C } \right)^u \;
{1 \over 1-2 u} \;.
\label{vertex:u=1/2} 
\end{equation}

There is also a vertex correction involving a Coulomb gluon with $A_0$
vertices and a ${\bf D}^4/(8m^3)$ correction in either the
quark or antiquark propagator. The resulting renormalization
coefficients are 
\begin{eqnarray} 
\delta\tilde{Z}_{V,b}=\delta\tilde{Z}_{P,b}=
4\pi i C_F \left( { \mu^2 \over e^C } \right)^u\; 
\int_\lambda^{\rm dim} {d^4 l \over (2\pi)^4} \;
&&{\mbox{\boldmath $l$}^4 \over 4m^3 } \; 
{1 \over (l_0 -  \mbox{\boldmath $l$}^2/(2m) + i \epsilon  )^2 } \;
{1 \over -l_0 - \mbox{\boldmath $l$}^2/(2m) +i \epsilon } \;\nonumber\\
&& \times {1  \over  \mbox{\boldmath $l$}^2 \, (-l^2 - i\epsilon)^{ u} }  \;,
\label{D^4} 
\end{eqnarray} 
where we have doubled the contribution with a ${\bf D}^4/(8m^3)$
insertion in the quark propagator in order to account for the equal
contribution with a ${\bf D}^4/(8m^3)$ insertion in the antiquark
propagator. Integrating over $l_0$ by closing the contour in the upper
half-plane, we obtain 
\begin{equation}
\delta\tilde{Z}_{V,b}=\delta\tilde{Z}_{P,b}\sim
{4\pi \over 3m}  \left( { \mu^2 \over e^C } \right)^u\; 
\int_\lambda^{\rm dim} {d^3 \mbox{\boldmath $l$} \over (2\pi)^3} \;
    {1  \over  \mbox{\boldmath $l$}^{2+2u} }\,,
\label{D^4:3q} 
\end{equation}
where, as usual, we have dropped the cut contribution, which does not
yield a pole at $u=1/2$. Carrying out the $\mbox{\boldmath $l$}$
integration, we identify the $u= {1\over 2}$ renormalon: 
\begin{equation}
\delta\tilde{Z}_{V,b}=\delta\tilde{Z}_{P,b}\sim
-{2\over 3\pi m} \;  
\left( { \mu^2 \over e^C } \right)^u \;
{1 \over 1-2 u} \;. 
\label{D^4:u=1/2} 
\end{equation}

Next we consider a vertex correction involving a Coulomb gluon with one
$A_0$ vertex and one ${\bf D}\cdot{\bf E}$ vertex. It yields the
renormalization coefficients 
\begin{eqnarray}
\delta\tilde{Z}_{V,c}=\delta\tilde{Z}_{P,c}= 
4\pi i C_F \left( { \mu^2 \over e^C } \right)^u\; 
\int_\lambda^{\rm dim} {d^4 l \over (2\pi)^4} \;
&&{\mbox{\boldmath $l$}^2 \over 4m^2 } \; 
{1 \over l_0 - \mbox{\boldmath $l$}^2 /(2m)+ i \epsilon }  \;
{1 \over -l_0 - \mbox{\boldmath $l$}^2 /(2m)+ i \epsilon } \;\nonumber\\
&&\times{1  \over \mbox{\boldmath $l$}^2 \, (-l^2 - i\epsilon)^{ u} }  \;,
\label{D.E} 
\end{eqnarray} 
where we have taken into account the two diagrams obtained by
interchanging the $A_0$ vertex and the ${\bf D}\cdot{\bf E}$ vertex.
This expression differs from (\ref{D^2}) by an overall factor $1/4$.
Thus from (\ref{D^2:u=1/2}), we obtain the contribution to the $u=1/2$
renormalon: 
\begin{equation}
\delta\tilde{Z}_{V,c}=\delta\tilde{Z}_{P,c}\sim
{2 \over 3\pi m} \;  
\left( { \mu^2 \over e^C } \right)^u \;
{1 \over 1-2 u} \;.
\label{D.E:u=1/2} 
\end{equation}

Finally, we consider the vertex correction involving a transverse gluon
with two $\mbox{\boldmath $\sigma$}\cdot{\bf B}$ vertices. These
vertices violate the spin symmetry, and, so, the vertex correction
depends on the spin structure of the original matrix element. The
resulting renormalization coefficients are
\begin{eqnarray} 
\left.\begin{array}{r}
\delta\tilde{Z}_{V,d}\\
\delta\tilde{Z}_{P,d}
\end{array}\right\}\Sigma
=-4\pi i C_F \left( { \mu^2 \over e^C } \right)^u\; 
\int_\lambda^{\rm dim} {d^4 l \over (2\pi)^4} \;
&&{(\mbox{\boldmath $\sigma$}\times \mbox{\boldmath $l$}  )_i \over 2m} \; 
{1 \over l_0 - \mbox{\boldmath $l$}^2/(2m)+ i \epsilon }  \; \Sigma\; 
{1 \over -l_0 - \mbox{\boldmath $l$}^2/(2m)+ i \epsilon } \;\nonumber\\
&&\times{(\mbox{\boldmath $\sigma$}\times \mbox{\boldmath $l$}  )_j \over 2m}\; 
{\delta_{ij}-l_il_j/\mbox{\boldmath $l$}^2 \over (-l^2 - i\epsilon)^{ 1+u} }\;,
\label{B.sigma} 
\end{eqnarray} 
where
\begin{equation}
\Sigma=\cases{\mbox{\boldmath$\sigma$}&\cr 1.&\cr}
\end{equation}
Integrating over $l_0$ by closing the contour in the upper half-plane
and simplifying the integrand by averaging over the angles of
$\mbox{\boldmath $l$}$, we obtain 
\begin{equation}
\left.\begin{array}{r}
\delta\tilde{Z}_{V,d}\\
\delta\tilde{Z}_{P,d}
\end{array}\right\}\Sigma
\sim {8\pi \over 9m} \, {\sigma_i \Sigma \sigma_i}\,
 \left( { \mu^2 \over e^C } \right)^u\; 
\int_\lambda^{\rm dim} {d^3 \mbox{\boldmath $l$} \over (2\pi)^3} \;
    {1  \over  \mbox{\boldmath $l$}^{2+2u} }\;, 
\label{B.sigma:3q} 
\end{equation}
where we have discarded the cut contribution, which does not contain a
pole at $u=1/2$.  Carrying out the $\mbox{\boldmath $l$}$ integration, we can
identify the $u= {1\over 2}$ renormalon: 
\begin{equation}
\left.\begin{array}{r}
\delta\tilde{Z}_{V,d}\\
\delta\tilde{Z}_{P,d}
\end{array}\right\}
\sim {1\over \pi m}\;\left( { \mu^2 \over e^C }\right)^u \; 
{1 \over 1-2u}\; 
\cases{+4/9 &\cr -4/3\; .&\cr}
\label{B.sigma:u=1/2} 
\end{equation}
It may be seen, from power-counting arguments, that the diagrams we have 
discussed so far are the only ones that contribute a pole at $u=1/2$.
Adding the contributions in Eqs.~(\ref{self:u=1/2}),
(\ref{vertex:u=1/2}), (\ref{D^4:u=1/2}), (\ref{D.E:u=1/2}), and
(\ref{B.sigma:u=1/2}) we obtain
\begin{equation}
\left.\begin{array}{r}
\delta\tilde{Z}_{V}\\
\delta\tilde{Z}_{P}
\end{array}\right\}
\sim {1\over \pi m}\;\left( { \mu^2 \over e^C }\right)^u \; 
{1 \over 1-2u}\; 
\cases{7/9&\cr -1\;.&\cr}
\label{renorm-total} 
\end{equation}
Therefore, from Eq.~(\ref{ambigu:2}), we see that the ambiguities in the
operators ${\cal K}_\psi$ and ${\cal K}_\eta$ are 
\begin{mathletters}
\begin{eqnarray}
\Delta \langle {\cal K}_{\psi } \rangle &\;=\;& -K{7 \over 9 \beta_0 }
 {\Lambda_{\rm QCD} \over e^{C/2} m} \langle {\cal K}_{\psi} \rangle \; ,\\
\Delta \langle {\cal K}_{\eta }\rangle  &\;=\;&K{1 \over  \beta_0 }
 {\Lambda_{\rm QCD} \over e^{C/2} m} \langle {\cal K}_{\eta} \rangle \; .
\end{eqnarray}
\label{1-ambig}%
\end{mathletters}

The results of our calculations of the $u=1/2$ renormalon in the 
operator matrix elements for S-wave charmonium decay are summarized in 
Table~\ref{table:results}.

We note that in NRQCD, in contrast with HQET and with the light-cone
expansion, the renormalon ambiguities scale differently from the matrix
elements themselves. For example, the $u=1/2$ renormalons computed in
this section have ambiguities that scale as $\Lambda_{\rm QCD}/m$,
relative to the matrix elements of lowest order in $v$, but the matrix
elements themselves scale relatively as $v^0$ or $v^2$. The differences
in scaling behavior between the renormalons and the matrix elements
arise because the dynamics of heavy-quarkonium system depends on the
scales $mv$ and $mv^2$, rather than the scale $\Lambda_{\rm QCD}$.

In Ref.~\cite{braaten-chen}, Braaten and Chen studied the large-order
asymptotic behavior of the perturbation series for short-distance
coefficients in the NRQCD factorization formulas for the decays $J/\psi
\to e^+e^-$ and $\eta_c \to \gamma\gamma$. They calculated the Borel
transforms of the short-distance coefficients of the leading (in $v^2$)
and first subleading matrix elements. By requiring that the renormalon
ambiguities in the short-distance coefficients be cancelled by the
renormalon ambiguities in the NRQCD matrix elements, they deduced the
$u=1/2$ renormalon ambiguities in the matrix elements.\footnote{The matrix 
elements of Braaten and Chen differ from the ones that we consider here by a 
factor $\protect\sqrt{2 M_H}$, where $M_H$ is the quarkonium mass.  However, 
$M_H$ is ambiguity free, so this change in normalization has no effect 
on the sizes of the ambiguities relative to the lowest-order-in-$v$
matrix elements.} Comparing the results obtained in
Ref.\cite{braaten-chen} with the results of our direct calculation of
the matrix-element ambiguities in Eqs.~(\ref{D^2-ambig}) and
(\ref{1-ambig}), we find agreement. Thus, the general principle of the
cancellation of renormalon ambiguities between the short-distance
coefficients and the operator matrix elements is supported by this
specific example in context of the NRQCD factorization formalism. 


\section{Renormalons in the Gremm-Kapustin Relation and the Quark Mass}
\label{sec:gremm-kapustin}

In this section, we discuss the renormalon content of the matrix
elements that appear in the Gremm-Kapustin relation
\cite{gremm-kapustin} and the renormalon content of various definitions
of the heavy-quark mass. 

The Gremm-Kapustin relation is an equality between NRQCD matrix elements
that relates the heavy-quark kinetic-energy operator to the energy of
the quarkonium meson. It is correct at leading order in $v$.  One form
of the Gremm-Kapustin relation follows immediately from the
leading-order equations of motion of the NRQCD Lagrangian. In the case
of the $J/\psi$ state, for example, we have 
\begin{equation}
\langle 0|\chi^\dagger {\bf D}^2
\mbox{\boldmath $\sigma$}\psi|J/\psi\rangle
=-m\langle 0|i\partial_0(\chi^\dagger \mbox{\boldmath 
$\sigma$}\psi)|J/\psi\rangle.
\label{gremm-kapustin-1}
\end{equation}
There are analogous relations for states with different spin and 
orbital-angular-momentum quantum numbers. 

The $u=1/2$ renormalon in the matrix element on the left side of
Eq.~(\ref{gremm-kapustin-1}) is proportional to the matrix element of
the lowest order operator $\chi^\dagger \mbox{\boldmath $\sigma$}\psi$.
The constant of proportionality is given by $m^2$ times
Eq.~(\ref{D^2:u=1/2}).  The right side of Eq.~(\ref{gremm-kapustin-1})
is proportional to the NRQCD energy of the quark-antiquark state 
$E_\psi$:
\begin{equation}
\langle 0|i\partial_0(\chi^\dagger \mbox{\boldmath 
$\sigma$}\psi)|J/\psi\rangle
=E_\psi\langle 0|\chi^\dagger \mbox{\boldmath 
$\sigma$}\psi|J/\psi\rangle.
\label{E_psi}
\end{equation}
The mixing of $i\partial_0(\chi^\dagger \mbox{\boldmath $\sigma$}\psi)$
into the lower-order operator $\chi^\dagger \mbox{\boldmath
$\sigma$}\psi$ is obtained by evaluating the QCD corrections to the
matrix element $\langle 0|i\partial_0(\chi^\dagger \mbox{\boldmath
$\sigma$}\psi)|c\bar c\rangle$, with the quark and antiquark taken on
shell and at zero external momentum.  Beyond tree level, the mass-shell
energy at zero momentum is shifted from zero to 
\begin{equation}
E_0=\Sigma(E,{\bf p})|_{E={\bf p}=0},
\label{E_0}
\end{equation}
where $\Sigma(E,{\bf p})$ is the quark (or antiquark)
self-energy (see Eq.~(\ref{self:Sigma})). This shift in the mass-shell
position yields a mixing of $i\partial_0(\chi^\dagger \mbox{\boldmath
$\sigma$})\psi$ into $\chi^\dagger \mbox{\boldmath $\sigma$}\psi$, with
a constant of proportionality $2E_0$. (The factor of two comes from the
sum of the equal quark and antiquark energy shifts.) Using
Eq.~(\ref{matrix-diff}), we can compute the renormalization coefficient
for the mixing of $2E_0$ into unity in passing from the cutoff regulator
to dimensional regularization. The contribution to the Borel transform
of that coefficient that comes from the graph containing a Coulomb gluon
with two $A_0$ vertices is given by 
\begin{eqnarray}
\delta\tilde{Z}_{V,\partial_0}=\delta\tilde{Z}_{E_\psi}
=2\delta\tilde{Z}_{E_0}
= 4 \pi i C_F \int_\lambda^{\rm dim}
{d^4 l \over (2\pi)^4} \;
&&\left[{1 \over l_0 -\mbox{\boldmath $l$}^2/2m + i\epsilon } \;
+{1 \over -l_0 -\mbox{\boldmath $l$}^2/2m + i\epsilon } \;\right]\nonumber\\
&&\times {1  \over \mbox{\boldmath $l$}^2 \, (-l^2 - i\epsilon)^{ u} }  \;,
\label{Z_E1} 
\end{eqnarray} 
where we have taken into account both the quark and antiquark
contributions, which has the effect of symmetrizing the integrand under
$l_0\rightarrow -l_0$. We have made explicit the fact that, to leading
order in $v$, $\delta\tilde{Z}_{V,\partial_0}=\delta\tilde{Z}_{E_\psi}$,
since, in Eq.~(\ref{E_psi}), any regulator dependence in $\langle
0|\chi^\dagger \mbox{\boldmath $\sigma$}\psi|J/\psi\rangle$ is
suppressed by the factor $E_\psi$, which is of order $v^2$. The
contribution in Eq.~(\ref{Z_E1}) is equal to $-m$ times the contribution 
in Eq.~(\ref{D^2}) and yields a pole at $u=1/2$: 
\begin{equation}
\delta\tilde{Z}_{V,\partial_0}=\delta\tilde{Z}_{E_\psi}
=2\delta\tilde{Z}_{E_0}
\sim -{8  \over 3 \pi} \;  
\left( { \mu^2 \over e^C } \right)^u \;
{1 \over 1-2 u} \;.
\label{Z_E2} 
\end{equation}
Hence, the left and right sides of Eq.~(\ref{gremm-kapustin-1}) have 
equal $u=1/2$-renormalon content.

Now, the NRQCD energy is related to the physical mass of the charmonium
state $M_{\psi}$ as 
\begin{equation}
M_{\psi}=E_\psi-2E_0+2m_{\rm pole},
\label{m_psi}
\end{equation}
where $m_{\rm pole}$ is the heavy-quark pole mass. In a theory without
confinement, such as Quantum Electrodynamics, both $m_{\rm pole}$ and
$E_0$ have nonperturbative definitions in terms of the poles in the
heavy-quark propagators in the full theory and the effective theory,
respectively. In a confining theory, such definitions are problematic;
only the linear combination $m_{\rm pole}-E_0$ has a nonperturbative
definition in terms of quarkonium matrix elements. Therefore, we
absorb $E_0$ into redefinition of $m_{\rm pole}$: 
\begin{equation}
m'_{\rm pole}=m_{\rm pole}-E_0,
\end{equation}
which implies that
\begin{equation}
2m'_{\rm pole}=M_\psi-E_\psi.
\label{m'_pole}
\end{equation}
Using this redefinition, we can rewrite the Gremm-Kapustin relation
(\ref{gremm-kapustin-1}) in the more conventional form 
\begin{equation}
\langle 0|\chi^\dagger {\bf D}^2\mbox{\boldmath $\sigma$}
\psi|J/\psi\rangle
=-m(M_\psi-2m'_{\rm pole})\langle 0|\chi^\dagger \mbox{\boldmath 
$\sigma$}\psi|J/\psi\rangle.
\label{gremm-kapustin-2}
\end{equation}

In dimensional regularization, $E_0$ vanishes and $m'_{\rm pole}$ is
equal to $m_{\rm pole}$. In the literature, $m_{\rm pole}$ and
$m'_{\rm pole}$ are used interchangeably. However, it is important to
bear in mind the differences between $m_{\rm pole}$ and $m'_{\rm
pole}$.\footnote{These remarks also are relevant in HQET. $\bar
\Lambda$, the heavy-light meson energy in HQET, is analogous to $E_\psi$
in NRQCD. In HQET $m'_{\rm pole}=M_{\rm meson}-\bar\Lambda$, where
$M_{\rm meson}$ is the physical meson mass.} In a confining theory,
$m_{\rm pole}$ is defined only in perturbation theory, whereas $m'_{\rm
pole}$ is defined, through Eq.~(\ref{m'_pole}) and Eq.~(\ref{E_psi}), in
terms of operator matrix elements. In a cutoff scheme, $m'_{\rm pole}$
exhibits a dependence, through $E_\psi$ on the UV cutoff, diverging as
its first power. (In a dimensional-regularization scheme, this power
dependence of $m'_{\rm pole}$ on the cutoff disappears.) $m'_{\rm pole}$
also depends on the effective theory in which it is defined. In
contrast, $m_{\rm pole}$ is independent of the cutoff and the effective
theory. This is true, by definition, in a nonconfining theory, where
$m_{\rm pole}$ is the physical mass, and it is enforced in perturbation
theory in a confining theory. 

In Eq.~(\ref{m_psi}), UV renormalons are absent in $E_0$ and $E_\psi$ in
cutoff regularization and cancel, by Eq.~(\ref{Z_E2}), in dimensional
regularization. Since $E_0$ vanishes in dimensional regularization,
the IR renormalons in $E_0$ must be equal in magnitude and opposite in
sign to the UV renormalons.\footnote{This is an example of the
replacement of IR renormalons with UV renormalons through the
manipulation of a quantity that vanishes in dimensional
regularization.}  $E_\psi$, being an operator matrix element, has no IR
renormalons. Therefore, we conclude that $m_{\rm pole}$ has IR
renormalon ambiguities, and that near $u=1/2$ 
\begin{equation}
\tilde{m}_{\rm pole}\sim {4  \over 3 \pi } \;  
\left( { \mu^2 \over e^C } \right)^u \;
{1 \over 1-2 u} \;.
\label{Z_pole} 
\end{equation}
Of course, we expect $m_{\rm pole}$ to contain IR renormalon
ambiguities, since the counterterm for on-shell mass renormalization
involves a loop integration down to zero momentum. It is important to
recognize that $m_{\rm pole}$ is neither a short-distance coefficient
nor an operator matrix element. Rather, $m_{\rm pole}$ is a quantity
that is defined perturbatively in terms of the pole in the heavy-quark
propagator. Consequently, $m_{\rm pole}$ is factorization-scheme
independent, and there are renormalon ambiguities in $m_{\rm pole}$,
even if one employs a cutoff factorization scheme. Of course, one could
absorb renormalon ambiguities in $m_{\rm pole}$ into cutoff matrix
elements of operators in NRQCD. However, the residual renormalon-free
short-distance coefficient would not be equal to the position of the
pole in the heavy-quark propagator. 

We remark that, in a dimensional-regularization factorization 
scheme, $m_{\rm pole}=m'_{\rm pole}$, and, so, 
\begin{equation}
\tilde{m}'_{\rm pole}\sim {4  \over 3 \pi } \;  
\left( { \mu^2 \over e^C } \right)^u \;
{1 \over 1-2 u} \;.
\end{equation} 

As with $m_{\rm pole}$, $m_{\overline{MS}}$ is neither a
short-distance coefficient nor an operator matrix element.
$m_{\overline{MS}}$ is related to the bare mass $m_0$ of full QCD as 
\begin{equation}
m_{\overline{MS}}=m_0+\delta m_{\overline{MS}}.
\label{m-msbar}
\end{equation}
$m_0$ is a parameter of full QCD and is independent of any ambiguities
associated with factorization; $\delta m_{\overline{MS}}$ consists of
poles in $\epsilon$ and associated constants, which contain no
ambiguities. Hence, in contrast with $m_{\rm pole}$, $m_{\overline{MS}}$
contains no ambiguities. 

Now we wish to demonstrate that our results for the ambiguities in
$E_\psi$, $E_0$ and $m_{\rm pole}$ are consistent with the fact that
$m_{\overline{MS}}$ has no ambiguities. First we re-express $m_{\rm
pole}$ in terms of the heavy-quark $\overline{MS}$ mass
$m_{\overline{MS}}$: 
\begin{equation}
\Delta m=m_{\rm pole}-m_{\overline{MS}}
=\Sigma^{\rm QCD}(p)|_{p\cdot\gamma=m_{\rm pole}}-\delta m_{\overline 
{MS}},
\end{equation}
where $\Sigma^{\rm QCD}(p)$ is the heavy-quark self-energy correction in
full QCD and $\delta m_{\overline{MS}}$ is the usual $\overline {MS}$
mass counterterm [$\delta m_{\overline{MS}}=(3C_F\alpha_s m/4\pi)
(1/\epsilon-\gamma_E+\ln 4\pi)$ in order $\alpha_s$]. Thus, we can
rewrite Eq.~(\ref{m_psi}) as 
\begin{equation}
E_\psi=M_\psi-2m_{\overline{MS}}+(2E_0-2\Delta m).
\label{E_psi-MS-bar}
\end{equation}
Since the expressions for the full QCD heavy-quark energy shift
$\Sigma^{\rm QCD}(p)|_{p\cdot\gamma=m_{\rm pole}}$ and the NRQCD energy
shift $E_0$ [Eq.~(\ref{E_0})] have, by construction, identical behavior
at small loop momentum, their difference $\Delta m-E_0$ has no support
at small loop momentum. Therefore, $\Delta m-E_0$ contains no IR
renormalon ambiguities. $E_\psi$, being an operator matrix element, has
no IR renormalon ambiguities. Furthermore, $\Delta m$ is free of UV
renormalon ambiguities because the counterterm $\delta m_{\overline
{MS}}$ removes the poles at $u=0$, which correspond to the UV
renormalons for a logarithmically divergent quantity such as
$\Sigma^{\rm QCD}$. We also see, from Eq.~(\ref{Z_E1}), that the quantity
$E_\psi-2E_0$ is free of UV renormalon ambiguities. Thus, in
Eq.~(\ref{E_psi-MS-bar}), $m_{\overline{MS}}$ must be free of both UV
and IR renormalon ambiguities. 

We note that the absence of renormalon ambiguities in $m_{\overline{MS}}$ 
implies that Eq.~(\ref{Z_pole}) is in agreement with the standard result
\cite{BB94} for the ambiguity of $m_{\rm pole}$ relative to
$m_{\overline{MS}}$.

\section{summary and discussion} 
\label{sec:conclusion}

Factorization formalisms for QCD are of great computational utility
because they allow one to separate short-distance, perturbative physics
from long-distance, nonperturbative effects. However, the perturbation
series for the short-distance coefficients in QCD are, at best,
asymptotic. Therefore, perturbative calculations of QCD processes are
ultimately limited in precision. 

One reason for the nonconvergence of the perturbation series is that,
even in the case of ``IR safe'' quantities, loop integrations typically
extend down to zero momentum. In that region, the running coupling
becomes large and perturbation theory fails. In the bubble-chain model
(described in Sec.~\ref{sec:renormalons}), the coefficients in the
perturbation series grow as the factorial of the order in $\alpha_s$,
leading to renormalon singularities in the Borel transform and to
ambiguities in the Borel summation of the series. 

By employing a cutoff factorization scheme, in which loop momenta
never become small, one can eliminate renormalon ambiguities from the
perturbation series for the short-distance coefficients. We wish to
emphasize that the renormalon ambiguities are emblematic of other
nonperturbative effects that could arise in the low-momentum,
long-distance region and which are not, as yet, well-understood. A cutoff
factorization scheme excludes this nonperturbative region from the
coefficient functions. In a dimensional-regularization factorization
scheme, one integrates all IR-finite terms in the integrand down to zero
momentum. Hence, dimensionally-regulated short-distance coefficients
contain renormalon ambiguities. 

Physical observables are free of ambiguities, in the sense described in
Section~\ref{sec:introduction}. This implies that ambiguities in the
factorized expression for a physical observable must cancel between the
short-distance coefficients and the operator matrix elements (provided
that the factorization procedure and the underlying effective field
theory are sufficiently accurate as an expansion in inverse powers of
the large scale in the process). That is, the ambiguities are merely
artifacts of the factorization procedure. 

The presence of ambiguities in the dimensionally-regulated
short-distance coefficients implies that at least some of the
dimensionally-regulated matrix elements contain ambiguities. Since
operator matrix elements are determined completely, in principal, in
terms of short-distance coefficients and physical observables, the
absence of ambiguities in the cutoff short-distance coefficients implies
that the cutoff matrix elements are free of ambiguities. Lattice
regularization is one example of a cutoff regulator and, indeed, the QCD
operator matrix elements are defined unambiguously in lattice
regularization.

The cancellation of ambiguities in a physical observable holds, by
construction, if one infers the matrix elements through a comparison of
theoretical expressions with observables. If one determines the matrix
elements through a calculation in an underlying effective field theory,
as we have done in this paper, then the cancellation of ambiguities
requires that the effective theory reproduce the low-momentum behavior
of the original theory. 

We stress that, in computing QCD corrections to the
dimensionally-regulated matrix elements, it is essential to adopt a
convention that is consistent with the standard computations of the
short-distance coefficients, in which IR-finite expressions are
integrated down to zero loop momentum.  This convention, which is
described in detail in Sec.~\ref{sec:dim-reg-prescrip}, involves writing
the loop corrections as linear combinations of operators in the
underlying effective theory and then expanding the loop integrals in
powers of the loop momentum divided by the large scale in the
factorization formalism. The net effect is to remove from the matrix
elements the power UV divergences that correspond to the IR-finite terms
in the short-distance coefficients. 

In this paper, we have presented a method for computing the renormalon 
ambiguities in dimensionally-regulated matrix elements. The method 
exploits the fact that cutoff matrix elements are unambiguous, 
which implies that one can compute the ambiguities in 
dimensionally-regulated matrix elements by computing the differences 
between the cutoff and dimensionally-regulated matrix elements. 
These differences are short-distance quantities and, hence, can be 
computed in perturbation theory.  

We have used our method to compute the $u=1/2$ renormalon ambiguities in
the matrix elements that appear in the NRQCD factorization expressions
for the annihilation decays of $S$-wave heavy quarkonium at the leading
and first subleading orders in $v$. On comparison with the results of
Braaten and Chen \cite{braaten-chen} for the renormalon ambiguities in
the corresponding short-distance coefficients, we find that the
ambiguities cancel in the physical decay rates, as expected. The
Gremm-Kapustin relation between operator matrix elements is also
consistent with the ambiguities that we find. Our result for the
ambiguity in $m_{\rm pole}$ is consistent with the ambiguity in the
expression for $m_{\rm pole}$ in terms of $m_{\overline{MS}}$ given in
Ref.~\cite{BBB}. 

In analyzing renormalon ambiguities, it is important to bear in mind
that $m_{\rm pole}$ and $m_{\overline{MS}}$ are neither short-distance
coefficients nor operator matrix elements. $m_{\overline{MS}}$ is equal
to the sum of the bare mass $m_0$ of full QCD and the $\overline{MS}$
mass counterterm, both of which are free of ambiguities and independent
of the factorization scheme. $m_{\rm pole}$ is a quantity that is
defined in perturbation theory by the on-shell renormalization condition
for a heavy quark. Because the on-shell renormalization counterterms
involve integrations down to zero momentum, $m_{\rm pole}$ is ambiguous.
$m_{\rm pole}$ is also independent of the factorization scheme.

The accounting of renormalon ambiguities that we have described in this
paper can be obscured if one writes operator matrix elements in terms of
$m_{\rm pole}$. For example, the matrix element for the energy of the
quarkonium state in NRQCD can be written [Eq.~(\ref{m_psi})] in terms of
the physical quarkonium mass $M_\psi$, $m_{\rm pole}$, and the NRQCD
energy shift $E_0$. Since $E_0$ vanishes in dimensional regularization,
one can use this relation to trade ambiguities in the NRQCD
quarkonium-energy matrix element for ambiguities in $m_{\rm
pole}$.\footnote{This relation, with $E_0$ set to zero, is often taken
as a definition of the pole mass. However, one can neglect $E_0$ only in
a dimensional-regularization factorization scheme.} The analogue of this
procedure for HQET is frequently employed in discussions of heavy-light
mesons. As a further step, one can remove the ambiguity in $m_{\rm
pole}$ and the corresponding ambiguity in the short-distance
coefficients by writing $m_{\rm pole}$ in terms of $m_{\overline{MS}}$. 

The renormalization procedure for full QCD can introduce additional IR
renormalons into the short-distance coefficients if the renormalization
counterterms themselves contain IR renormalons. This occurs whenever the
counterterms involve loop integrations down to zero momentum, as is the 
case in on-shell renormalization. The ambiguities associated with these
additional renormalons are not cancelled by ambiguities in the operator
matrix elements, but, rather, by ambiguities in the parameters of full
QCD. For example, because $m_{\rm pole}$ is ambiguous, one introduces
additional ambiguities into the short-distance coefficients if one
expresses them in terms of $m_{\rm pole}$, rather than, say,
$m_{\overline{MS}}$. The additional ambiguities in the short-distance
coefficients are cancelled by the ambiguities in $m_{\rm pole}$. 

In practice, one determines dimensionally-regulated matrix elements
either by comparing perturbative expressions for physical observables
with experiment or by computing the perturbative relations between
dimensionally-regulated matrix elements and some reference cutoff
matrix elements, such as lattice matrix elements. In both methods, the
ambiguities in the perturbation series make the determinations of the
matrix elements ambiguous. Hence, in order to describe a
dimensionally-regulated matrix element, one must give not only its
value, but also the order in $\alpha_s$ in which it was determined. At
low orders in $\alpha_s$, this last specification may not be important
numerically, but at high orders, when the factorial growth of the
perturbation series dominates, it is essential. 

Finally, let us discuss the relative strengths and weaknesses of the
dimensional-regularization and cutoff factorization schemes.
Dimensionally-regulated short-distance coefficients can be computed
relatively efficiently, but the computation of, for example,
lattice-regulated short-distance coefficients, is much more complicated.
The dimensionally-regulated short-distance coefficients and matrix
elements are ambiguous, whereas the cutoff short-distance coefficients
and matrix elements are not.  Of course, the presence of renormalon
ambiguities in the dimensional-regularization scheme is not a fatal
flaw: In physical observables, the factorial growth in the
dimensionally-regulated short-distance coefficients cancels against a
similar growth in the dimensionally-regulated matrix elements. However,
in order to achieve this cancellation, one must work to the same order
in $\alpha_s$ in the short-distance coefficients and the matrix
elements.  In addition, it is essential to work to sufficient accuracy
in the small scale of the underlying effective field theory
\cite{LMS94}. In practice, one cannot achieve a complete cancellation,
but one can systematically reduce the renormalon ambiguities by
introducing more operator matrix elements and/or including more terms in
the effective action. In general, cutoff matrix elements are very
sensitive to the cutoff (factorization scale), typically scaling as a
power of the cutoff. This sensitivity to the cutoff is cancelled by a
similar dependence in the short-distance coefficients, but the
cancellation is imperfect at any finite order in perturbation theory. A
related difficulty is that the cutoff matrix elements exhibit their
nominal sizes only if the cutoff is chosen to be of the order of the
small scale in the calculation. In contrast, dimensionally-regulated
($MS$) matrix elements have no power dependence on the cutoff.
Nevertheless, at high orders in $\alpha_s$, they must deviate from their
nominal sizes because of the factorial growth of the determining
perturbation series. 

Of course, in the end, the physical observables are independent of the
choice of factorization scheme, provided that one works to sufficient
accuracy in the small scale of the effective theory and to sufficient
and consistent accuracy in $\alpha_s$. However, it is important to bear
in mind that the properties of the dimensionally-regulated and cutoff
quantities are rather different and that the differences may be
significant in practical calculations. 


\acknowledgements 

We acknowledge valuable discussions with M.~Beneke, E.~Braaten,
G.P.~Lepage, E.~Kov\'acs, and G.~Sterman. This work was supported in
part by the U.S. Department of Energy, Division of High Energy Physics,
under Contract No.\ W-31-109-ENG-38 and under Grant No.\
DE-FG02-91-ER40684 .

\vfill \eject

\begin{table}
\begin{center}
\begin{tabular}{ll}
Diagrammatic Element & Feynman Rule \\
\hline
quark propagator 
& $i/[\pm p_0-{\bf p}^2/(2m)+i\epsilon]$\\
$A_0$ vertex 
& $\mp ig\,t_a$ \\
$\mbox{\boldmath $\nabla$}\cdot {\bf A}_i$ vertex 
& $ig(p_i+p'_i)\,t_a/(2m)$\\
${\bf A}_i\cdot {\bf A}_j$ seagull vertex 
& $-ig^2(\delta_{ij}\,t_bt_a+\hbox{perm})/(2m)$\\
$\mbox{\boldmath $\sigma$}\cdot {\bf B}$ spatial-gluon vertex 
& $g\epsilon_{ijk}l_{1j} \sigma_k \,t_a /(2m)$\\
${\bf D}\cdot {\bf E}$ temporal-gluon vertex 
&$\pm ig \mbox{\boldmath $l$}_1^2 \,t_a/(8m^2)$\\
${\bf D}\cdot {\bf E}$ spatial-gluon vertex
& $\mp igl_{1i}l_{10} \,t_a/(8m^2)$\\
${\bf D}\times {\bf E} \cdot \mbox{\boldmath
$\sigma$}$ temporal-gluon vertex
& $\pm g\epsilon_{ijk} p'_ip_j\sigma_k \,t_a/(4m^2)$\\
${\bf D}\times {\bf E} \cdot \mbox{\boldmath 
$\sigma$}$ spatial-gluon vertex 
& $\mp g \epsilon_{ijk}(p+p')_j\sigma_kl_{10} \,t_a/(8m^2)$\\
${\bf D}\times {\bf E} \cdot \mbox{\boldmath
$\sigma$}$ spatial-temporal seagull vertex
& $\pm g^2(\epsilon_{ijk} l_{1j}\sigma_k \,t_bt_a+\hbox{perm})/(4m^2)$\\
${\bf D}\times {\bf E} \cdot \mbox{\boldmath
$\sigma$}$ spatial-spatial seagull vertex
& $\pm g^2(\epsilon_{ijk}\sigma_k l_{10}\,t_bt_a+\hbox{perm})/(4m^2)$\\
${\bf D}^4/(8m^3)$ quark-propagator correction
& $i{\bf p}^4/(8m^3)$\\
${\bf D}^4/(8m^3)$ spatial-gluon vertex
&$-ig({\bf p}^2+{\bf p'}^2)(p+p')_i\ \,t_a/(8m^3)$\\
${\bf D}^4/(8m^3)$ spatial-gluon seagull vertex
&$ig^2\{[(2p'-l_2)_j (2p+l_1)_i+({\bf p}^2+{\bf p'}^2)\delta_{ij}]
\,t_bt_a$\\
&$\qquad\qquad +\hbox{perm}\}/(8m^3)$\\
${\bf D}^4/(8m^3)$ 3-spatial-gluon vertex
&$-ig^3\{[(2p'-l_3)_k\delta_{ij}+(2p+l_1)_i\delta_{kj}]
\,t_ct_bt_a$\\
&$\phantom{-}\qquad\qquad +\hbox{perm}\}/(8m^3)$\\
${\bf D}^4/(8m^3)$ 4-spatial-gluon vertex
&$ig^4(\delta_{ij}\delta_{km}\,t_dt_ct_bt_a+\hbox{perm})/(8m^3)$\\
\end{tabular}
\caption{\label{table:rules}
Feynman rules for quark and antiquark propagators and interaction
vertices in NRQCD. The upper (lower) signs correspond to quarks
(antiquarks). We have retained only the ``Abelian'' terms, i.e.,
those that contain no color-matrix commutators. Energy and momentum are
conserved at the vertices, and we observe the convention in which the
antiquark diagrammatic energy and momentum are the negatives of the
physical energy and momentum. $p$ is the incoming (diagrammatic) quark
or antiquark momentum, $p'$ is the outgoing (diagrammatic) quark or
antiquark momentum. The gluons have incoming momenta, spatial
polarization indices, and color indices $(l_1,i,a)$, $(l_2,j,b)$,
$(l_3,k,c)$, and $(l_4,m,d)$, respectively.  $T$ denotes an
$SU(3)$ color matrix in the fundamental representation, normalized to
${\rm tr}\, t_at_b=(1/2)\delta_{ab}$. ``Perm'' denotes all additional 
permutations obtained by interchanging the momentum, polarization, and 
color labels of one gluon with those of another gluon.
} 
\end{center}
\end{table}

\begin{table}
\begin{center}
\begin{tabular}{llll} 
Quantity &Description &Spin Triplet &Spin Singlet\\
\hline 
Correction to $\langle{\cal K}_D^2\rangle$:&&&\\
$\delta\tilde{Z}_{V,D^2}/\delta\tilde{Z}_{P,D^2}$&$A_0$-$A_0$ vertex 
correction&+8/3&+8/3\\
\hline
Corrections to $\langle{\cal K}\rangle$:&&&\\
$\delta\tilde{Z}_{\rm wf}$&Quark wave-function renormalization&$+2/3$&$+2/3$\\
$\delta\tilde{Z}_{V,a}/\delta\tilde{Z}_{P,a}$&$A_0$-$A_0$ vertex 
correction&$-1/3$&$-1/3$\\
$\delta\tilde{Z}_{V,b}/\delta\tilde{Z}_{P,b}$&$A_0$-$A_0$ vertex 
correction with ${\bf D}^4$ prop.\ insertion&$-2/3$&$-2/3$\\
$\delta\tilde{Z}_{V,c}/\delta\tilde{Z}_{P,c}$&$A_0$-${\bf D}\cdot{\bf E}
$ vertex correction&$+2/3$&$+2/3$\\
$\delta\tilde{Z}_{V,d}/\delta\tilde{Z}_{P,d}$&$\mbox{\boldmath 
$\sigma$}\cdot {\bf B}$-$\mbox{\boldmath $\sigma$}\cdot {\bf B}$ vertex 
correction&$+4/9$&$-4/3$\\
$\delta\tilde{Z}_{V}/\delta\tilde{Z}_{P}$&Total correction to 
$\langle{\cal K}\rangle$&$+7/9$&$-1$\\
\end{tabular}
\caption{\label{table:results} The $u=1/2$-renormalon contributions in
the mixing of the spin-triplet matrix and spin-singlet matrix elements
$\langle{\cal K}_D^2\rangle_V$ and $\langle{\cal K}_D^2\rangle_P$ into
the spin-triplet and spin-singlet matrix elements $\langle{\cal
K}\rangle_V$ and $\langle{\cal K}\rangle_P$, respectively, and the
$u=1/2$-renormalon contributions in the multiplicative renormalization
of the spin-triplet and spin-singlet matrix elements $\langle{\cal
K}\rangle_V$ and $\langle{\cal K}\rangle_P$. The values displayed are
the coefficients of $(1/\pi)(\mu^2/ e^C)^u (1-2u)^{-1}$ times the matrix
elements.} 
\end{center} 
\end{table} 
\end{document}